\documentclass[10pt,journal,compsoc]{IEEEtran}

\usepackage{scalerel}
\usepackage{tikz}
\usetikzlibrary{svg.path}
\definecolor{orcidlogocol}{HTML}{A6CE39}
\tikzset{
  orcidlogo/.pic={
    \fill[orcidlogocol] svg{M256,128c0,70.7-57.3,128-128,128C57.3,256,0,198.7,0,128C0,57.3,57.3,0,128,0C198.7,0,256,57.3,256,128z};
    \fill[white] svg{M86.3,186.2H70.9V79.1h15.4v48.4V186.2z}
                 svg{M108.9,79.1h41.6c39.6,0,57,28.3,57,53.6c0,27.5-21.5,53.6-56.8,53.6h-41.8V79.1z M124.3,172.4h24.5c34.9,0,42.9-26.5,42.9-39.7c0-21.5-13.7-39.7-43.7-39.7h-23.7V172.4z}
                 svg{M88.7,56.8c0,5.5-4.5,10.1-10.1,10.1c-5.6,0-10.1-4.6-10.1-10.1c0-5.6,4.5-10.1,10.1-10.1C84.2,46.7,88.7,51.3,88.7,56.8z};
  }
}
\newcommand\orcidicon[1]{\href{https://orcid.org/#1}{\mbox{\scalerel*{
\begin{tikzpicture}[yscale=-1,transform shape]
\pic{orcidlogo};
\end{tikzpicture}
}{|}}}}

\ifCLASSOPTIONcompsoc
  \usepackage[nocompress]{cite}
\else
  \usepackage{cite}
\fi

\usepackage{amsmath,amssymb,amsfonts}
\usepackage{graphicx}
\usepackage{textcomp}
\usepackage{subcaption}
\usepackage{multirow}
\usepackage{makecell}
\usepackage{amsthm}
\usepackage[T1]{fontenc}
\usepackage{enumitem}
\usepackage{comment}
\usepackage{multirow}
\usepackage{amsmath}

\usepackage{algorithm,algorithmicx,algpseudocode}
\usepackage{listings}
\usepackage{pifont}
\usepackage{booktabs}
\usepackage[hidelinks]{hyperref}
\usepackage{enumitem}
\setlist[itemize]{leftmargin=*}
\setlist[enumerate]{leftmargin=*}
\setlist{nolistsep}
\usepackage{amsmath}

\DeclareMathOperator*{\argmin}{arg\,min}

\def\BibTeX{{\rm B\kern-.05em{\sc i\kern-.025em b}\kern-.08em
    T\kern-.1667em\lower.7ex\hbox{E}\kern-.125emX}}

\graphicspath{ {images/} }
\newcommand{\vspacegraph}{\vspace{-0pt}}

\ifCLASSINFOpdf
\else
\fi

\usepackage{xcolor}

\definecolor{codegreen}{rgb}{0,0.6,0}
\definecolor{codegray}{rgb}{0.5,0.5,0.5}
\definecolor{codepurple}{rgb}{0.58,0,0.82}
\definecolor{backcolour}{rgb}{0.95,0.95,0.92}

\lstdefinestyle{mystyle}{
  backgroundcolor=\color{backcolour}, commentstyle=\color{codegreen},
  keywordstyle=\color{magenta},
  numberstyle=\tiny\color{codegray},
  stringstyle=\color{codepurple},
  basicstyle=\ttfamily\footnotesize,
  breakatwhitespace=false,         
  breaklines=true,                 
  captionpos=b,                    
  keepspaces=true,                 
  numbers=left,                    
  numbersep=5pt,                  
  showspaces=false,                
  showstringspaces=false,
  showtabs=false,                  
  tabsize=2
}

\definecolor{dkgreen}{rgb}{0,0.5,0}
\definecolor{ddkgreen}{rgb}{0,0.7,0}
\definecolor{gray}{rgb}{0.5,0.5,0.5}
\definecolor{mauve}{rgb}{0.58,0,0.82}

\definecolor{blue}{rgb}{0,0,1}

\begin{document}

%
\title{AutoDDL: Automatic Distributed Deep Learning with Near-Optimal Bandwidth Cost}
%



\author{Jinfan Chen, Shigang~Li~\orcidicon{0000-0003-0022-7865}, Ran Guo, Jinhui Yuan, Torsten~Hoefler~\orcidicon{0000-0002-1333-9797}
\IEEEcompsocitemizethanks{\IEEEcompsocthanksitem Jinfan Chen and Torsten Hoefler are with Department of Computer Science, ETH Zurich, Switzerland. E-mail: {jinfchen@student.ethz.cn, htor@inf.ethz.ch}

Shigang Li is with School of Computer Science, Beijing University of Posts and Telecommunications, China. E-mail: shigangli.cs@gmail.com

Ran Guo and Jinhui Yuan are with OneFlow, China.}
\thanks{(Corresponding Author: Shigang Li.) \\ Published in IEEE Transactions on Parallel and Distributed Systems (TPDS), Vol. 35, No. 8, Aug. 2024, DOI: \textcolor{blue}{\url{https://doi.org/10.1109/TPDS.2024.3397800}}}
}

%
%

\markboth{}%
{Shell \MakeLowercase{\textit{et al.}}: Bare Demo of IEEEtran.cls for Computer Society Journals}
%



\IEEEtitleabstractindextext{%
\begin{abstract}

Recent advances in deep learning are driven by the growing scale of computation, data, and models. However, efficiently training large-scale models on distributed systems requires an intricate combination of data, operator, and pipeline parallelism, which exerts heavy burden on machine learning practitioners. To this end, we propose AutoDDL, a distributed training framework that automatically explores and exploits new parallelization schemes with near-optimal bandwidth cost. AutoDDL facilitates the description and implementation of different schemes by utilizing OneFlow's \textit{Split}, \textit{Broadcast}, and \textit{Partial Sum} (SBP) abstraction. AutoDDL is equipped with an analytical performance model combined with a customized Coordinate Descent algorithm, which significantly reduces the scheme searching overhead. 
We conduct evaluations on Multi-Node-Single-GPU and Multi-Node-Multi-GPU machines using different models, including VGG and Transformer. Compared to the expert-optimized implementations, AutoDDL reduces the end-to-end training time by up to 31.1\% and 10\% for Transformer and up to 17.7\% and 71.5\% for VGG on the two parallel systems, respectively.
\end{abstract}

\begin{IEEEkeywords}
Distributed Deep Learning, Operator Parallelism, Data Parallelism, Pipeline Parallelism
\end{IEEEkeywords}}

\maketitle

\IEEEdisplaynontitleabstractindextext

%
\IEEEpeerreviewmaketitle

\section{Introduction}

The continuous success of deep learning hinges on the ability to handle quickly growing model sizes~\cite{scaling_law}. Recent breakthroughs in large language models, exemplified by ChatGPT (fine-tuned from GPT-3.5) and GPT-4~\cite{GPT4}, have made a profound impact on our daily lives. These advancements demonstrate the power of scaling up deep learning models, which helps to significantly increase the model accuracy.

\vspacegraph
\begin{figure}[!ht]
\centering
    \includegraphics[width=0.40\textwidth]{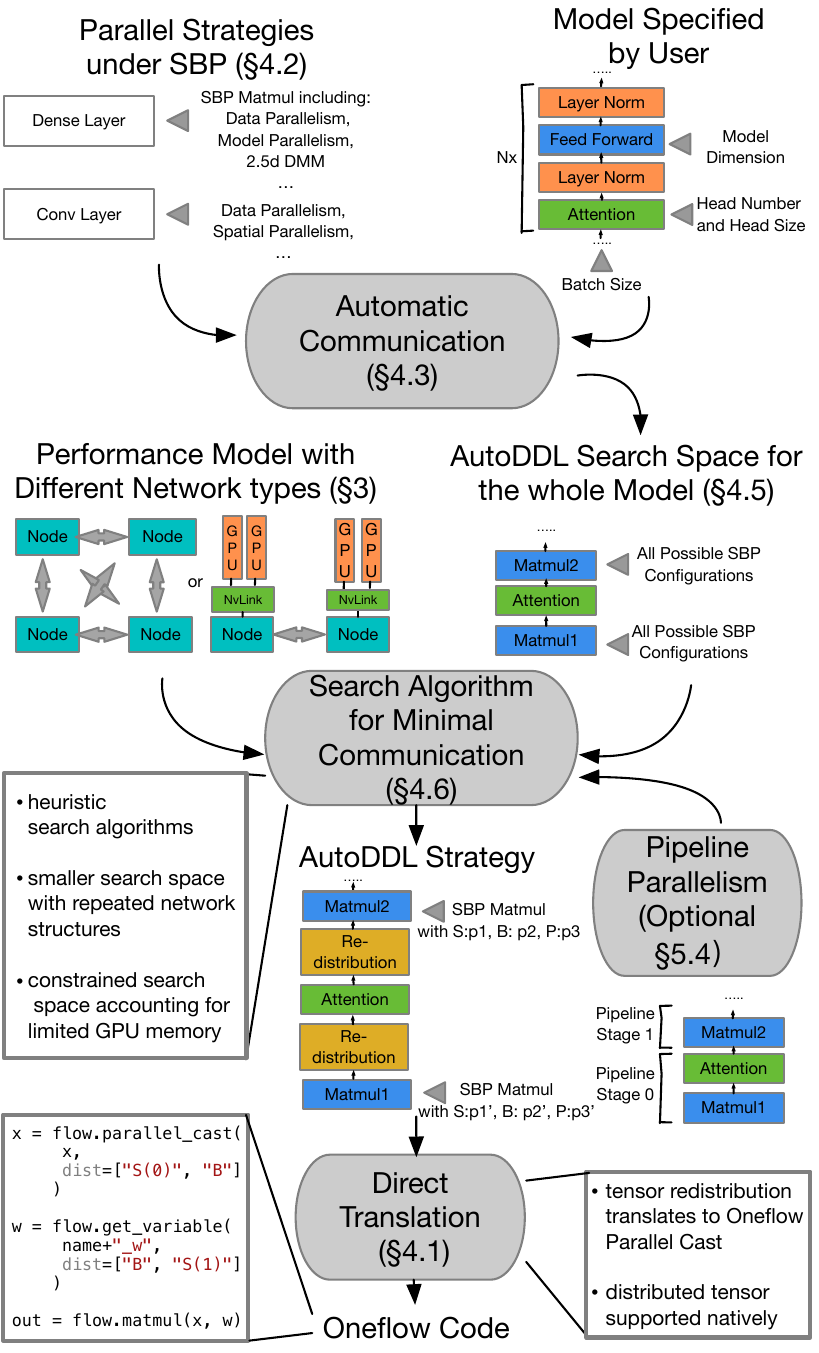}
    \vspace{-5pt}
\caption{The workflow of AutoDDL. The AutoDDL search process involves parsing a user-specified DNN, constructing a search space of distributed strategies in SBP according to an performance model, and automatically deploying an optimal strategy in Oneflow code.\label{fig:ada_framework}}
\vspace{-16pt}
\end{figure}

However, training these large models requires massive hardware resources. Therefore, distributed learning on parallel machines almost becomes the standard for large models. Although hardware vendors bring unprecedented AI computing power, training these large models is still time- and money-consuming. For instance, training a GPT-3 model with 175 billion parameters takes seven months on 512 V100 GPUs and costs millions of dollars~\cite{brown2020language}. The main performance bottleneck of distributed training is often the high communication cost~\cite{hoefler2022hammingmesh, 25d, optimus, ColossalAI}, which can significantly hinder the training efficiency.

To reduce the communication overhead of distributed training, it is imperative to design efficient parallelization schemes involving data~\cite{parameter_server, distdl, horovod, li2020taming}, operator~\cite{megatron, wierd_trick}, and pipeline~\cite{dapple, pipedream, gpip, pipemare, li2021chimera} parallelism. To describe parallelization strategies, GShard~\cite{gshard} provides abstractions for data and operator parallelism, where tensors can be \textit{split} or \textit{replicated} on different machines. Although there are several deep learning frameworks~\cite{alpa, flexflow, tofu} that utilize GShard's abstractions to automatically find efficient parallel strategies for deep neural networks (DNNs), the challenge of minimizing the communication overhead is still not solved. One of the main reasons is that the space of parallelization strategies is not adequately defined in existing work, which hinders these deep learning frameworks from exploring more communication-efficient schemes, such as 2.5D or 3D algorithms~\cite{Cosma, demmel, irony2004communication, Carma} for distributed matrix multiplication. Therefore, the parallelization strategies 
 discovered by the existing deep learning frameworks can be sub-optimal.

\vspacegraph
\begin{figure}[!t]
\centering
    \includegraphics[width=0.49\textwidth]{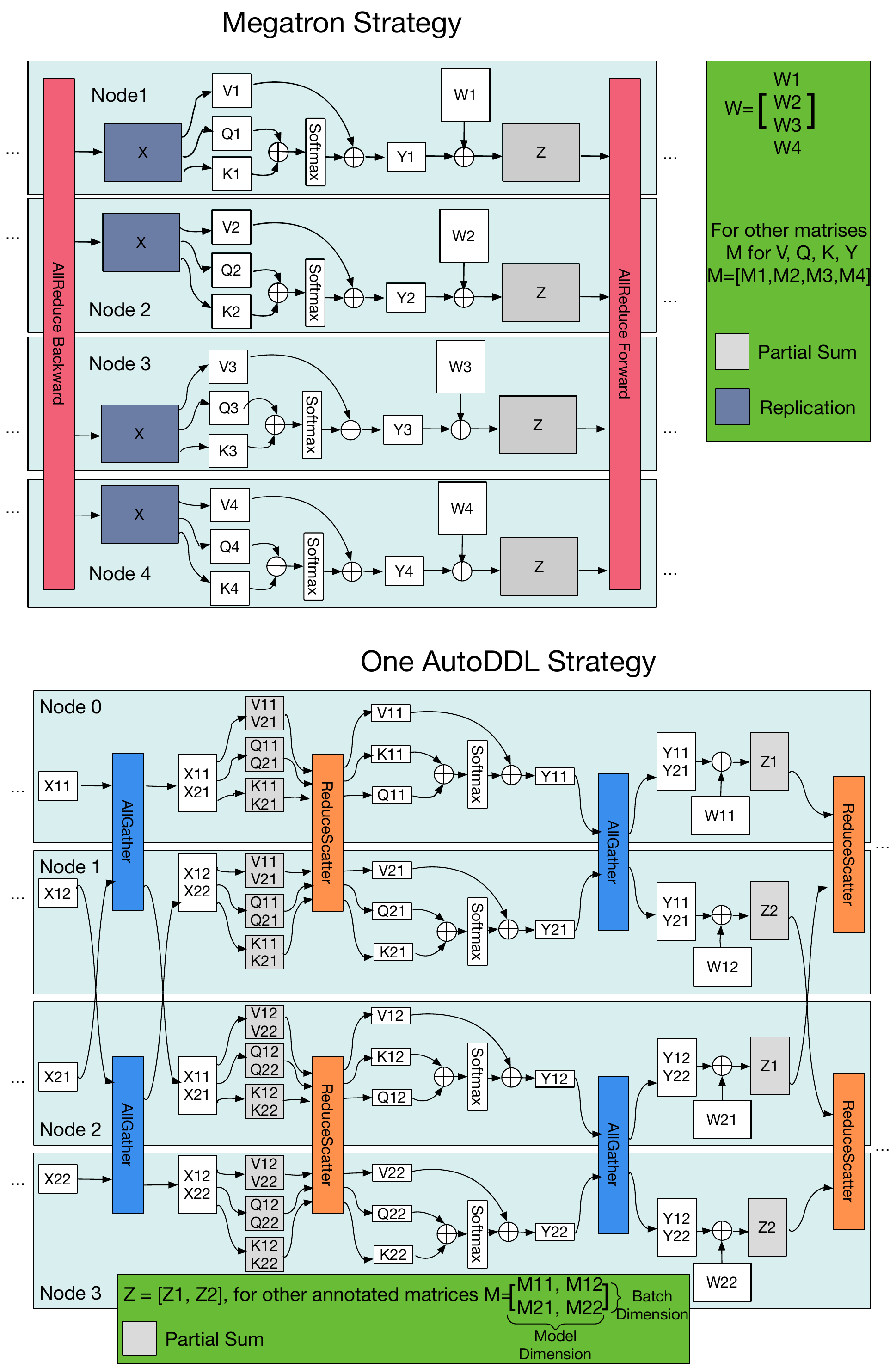}
    \vspace{-5pt} 
\caption{Comparison between Megatron-LM and a parallelization strategy explored by AutoDDL for the self-attention layer of Transformer on four compute nodes (We only present the operator parallelism in the figure and omit the data parallelism for brevity). \label{fig:simple_attention_example} \vspace{-8pt}}
\end{figure}

To this end, we propose an \textbf{\textit{Auto}}matic framework for \textbf{\textit{D}}istributed \textbf{\textit{D}}eep \textbf{\textit{L}}earning (AutoDDL). The workflow of AutoDDL is shown in Figure~\ref{fig:ada_framework}. AutoDDL expands the search space for parallelization strategies by leveraging the \textit{Split}, \textit{Broadcast} (the same as GShard's \textit{replicated}), and \textit{Partial Sum} (SBP) abstractions of the deep learning framework Oneflow~\cite{oneflow}. SBP extends GShard's abstraction by introducing a tensor state of \textit{Partial Sum}, which means that the final tensor can be obtained by performing an element-wise reduction (e.g., sum, max, etc.) operation over all the partial-sum tensors. Although GShard allows tensors to be split along the reduction dimension, it does not have an explicit partial sum state and the partial sums are immediately all-reduced in the output. In contrast, AutoDDL can keep the partial sums as is which may eliminate intermediate all-reductions. AutoDDL utilizes an analytical performance model combined with a Coordinate Descent-based search algorithm to discover the strategy with minimal communication cost, which leads to negligible searching overhead. Finally, the selected strategy is directly translated to the
parallel code of OneFlow, mainly including the tensor annotations for parallelization and the redistribution primitives to satisfy data layout requirements between continuous operators. Our study shows that the expanded search space enables AutoDDL to explore parallelization strategies with near-optimal communication costs. Furthermore, AutoDDL can adapt different parallel strategies to different neural network layers to minimize the end-to-end communication cost.


\vspacegraph
\begin{figure}[t]
\centering
    \includegraphics[width=0.49\textwidth]{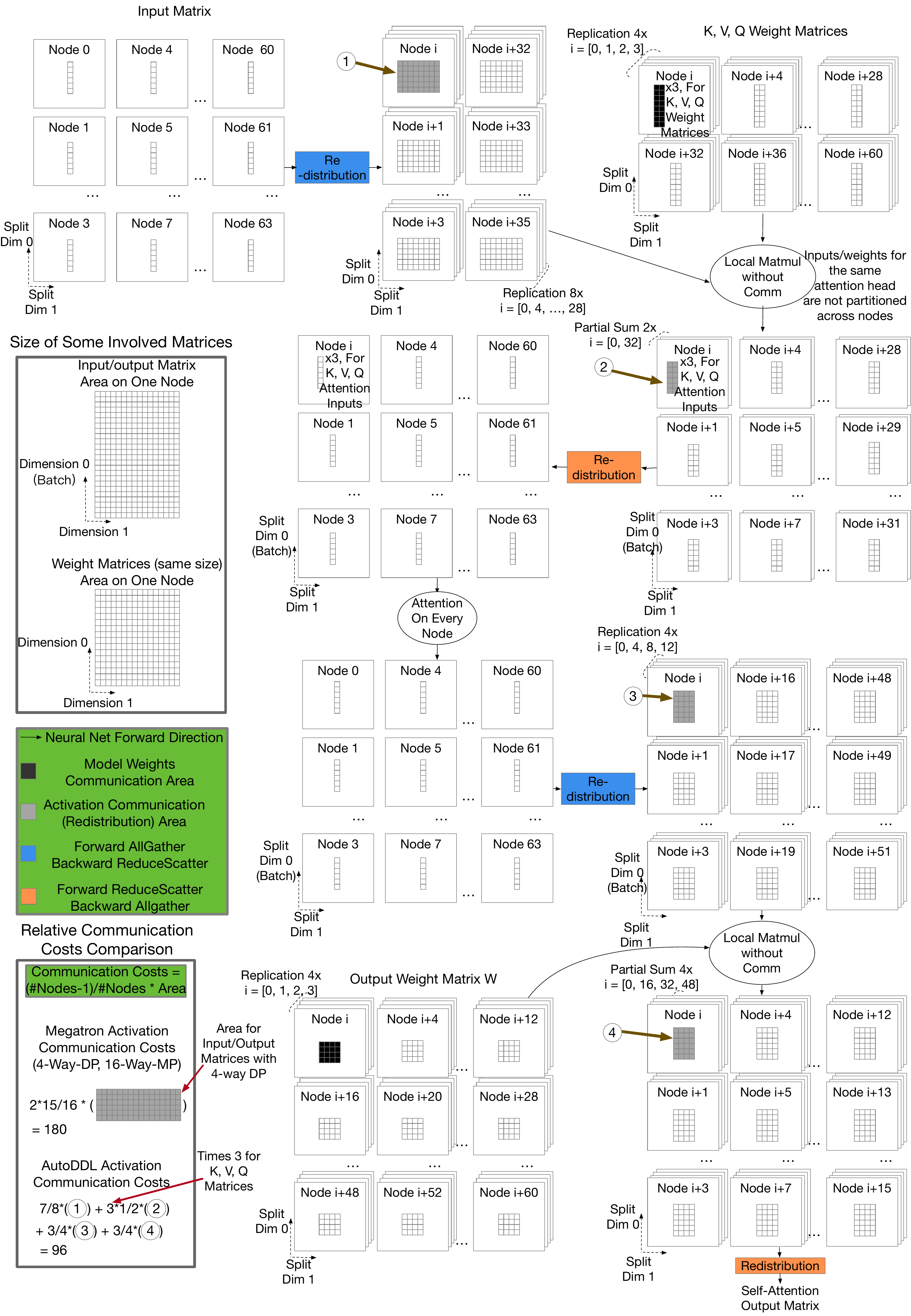}
    \vspace{-5pt} 
\caption{SBP parallel strategies for one Attention layer in our actual experiment using 64 nodes. The precise details of data distributions, communication patterns, and relative communication volumes are meticulously depicted.\label{fig:transformer_example} \vspace{-8pt}}
\end{figure}

Figure~\ref{fig:simple_attention_example} gives a comparison between Megatron-LM and a parallelization strategy explored by AutoDDL using the self-attention layer of Transformer as an example. As shown in the upper half of Figure~\ref{fig:simple_attention_example}, Megatron-LM replicates input X and splits model weights along one dimension for operator parallelism. For clarity, Fig.\ref{fig:simple_attention_example} depicts both AutoDDL and Megatron only in terms of operator parallelism. Introducing data parallelism to both strategies would result in the following complexity: For $p$ compute nodes and matrix size of $n^2$, Megatron-LM leads to communication and memory costs of $\mathcal{O}\left(\frac{n^2}{p^{1/2}} \right)$ when combined with data parallelism. In contrast, the strategy explored by AutoDDL splits the input X along the model dimension and splits the model weights along both dimensions. This allows AutoDDL to exploit more parallel dimensions, which results in more efficient parallelization schemes, such as 2.5D or 3D distributed matrix multiplication with $\Theta \left( \frac{n^2}{p^{2/3}} \right)$ communication and memory costs.

Figure~\ref{fig:transformer_example} shows an actual strategy selected by AutoDDL for an attention layer on 64 nodes. In this case, the model is large and requires more than 16 nodes to carry the model weights using model parallelism. The best strategy for Megatron is found to be 16-way model parallelism (MP) combined with 4-way data parallelism (DP). Consequently, Megatron's 16-way model parallelism needs to gather the intermediate activation across 16 nodes, with the mini-batch size divided by 4. While the strategy selected by AutoDDL has the same communication costs for weight gradient synchronization as Megatron (marked in black in Figure~\ref{fig:transformer_example}), the communication cost for gathering intermediate activation is nearly halved, as shown in the bottom left of Figure~\ref{fig:transformer_example}. The illustrated transformer has a model dimension of 8,192 and a batch size of 1,024 and a sequence length of 1,024. In the case of Megatron, the intermediate activation communication for each attention layer is 3.75 billion elements (15 GB if using float32). Furthermore, as illustrated in Figure~\ref{fig:transformer_example}, we present the concrete example of AutoDDL comparison. According to the 96/180 ratio of communication costs shown in Figure~\ref{fig:transformer_example}, AutoDDL's approach reduces communication overhead to 2 billion elements (8 GB if using float32), leading to a substantial cost saving.

\begin{table*}[t]
\makebox[\textwidth][c]{ 
  \begin{tabular}{ lcccc }
  \toprule
  
  & Supported Models & \makecell{Asymptotically Optimal \\ Communication Cost} & Strategy Search Cost & Usability \\
  \toprule
 Data Parallel~\cite{parameter_server} & Models fit in memory & No Guarantee & N/A & Easy \\
 \hline
 Operator Parallel~\cite{wierd_trick, GooglesNM} & All & No Guarantee & N/A & Hard \\
 \hline
 Megatron~\cite{megatron} & Transformers & No Guarantee  & N/A & Medium \\
 \hline
 Summa~\cite{summa} & Matmul & For DMM Communication & N/A & Medium \\
 \hline
 Flexflow~\cite{flexflow} & All & No Guarantee & High & Easy \\
 \hline
 Tofu~\cite{tofu} & partition-n-reduction & No Guarantee & High & Easy \\
 \hline
 Alpa~\cite{alpa} & All & No Guarantee & High & Easy \\
 \hline
 Colossal-AI~\cite{ColossalAI} & All & For DMM Communication & N/A & Easy but with Restrictions \\
 \hline
  AutoDDL & All & For DMM Communication & Low & Easy   \\

 \bottomrule
  \end{tabular}
  }
  \vspace{-5pt}
  \caption{Comparison of distributed deep learning methods. \vspace{-8pt}}
  \label{comparison}
\end{table*}

To discover a parallelization strategy with near-optimal communication cost, we built a performance model for AutoDDL, which accurately ranks the communication overhead of various parallelization strategies. Furthermore, AutoDDL extends the current search space for automatic distributed deep learning by introducing the \textit{Partial Sum} state, and therefore has the capacity to cover more competitive strategies in terms of communication cost in the space. However, the extended space leads to non-negligible overhead for exhaustive search even with the help of performance model. As such, we propose a customized coordinate descent algorithm for heuristic search, which significantly reduces the searching overhead and can find the best strategy even faster than the MCMC method used in Flexflow~\cite{flexflow}.


Our main contributions are:
\begin{itemize}

 \item AutoDDL supports a novel search space that can adapt different parallelization strategies to different operators according to the tensor shapes. The expanded search space is essential for exploring parallelization strategies with near-optimal
communication cost, which is not supported in existing frameworks.

 \item We establish a performance model that effectively guides the search for parallelization strategies with low communication overhead. To expedite the search process, we implement and evaluate several heuristic search algorithms, in which our customized coordinate descent algorithm performs best. Our algorithm takes only a few minutes on a personal laptop to discover a near-optimal parallelization strategy for large models.
 
 \item We integrate the deep learning framework OneFlow into the workflow of AutoDDL, which enables automatic implementation of various parallelization strategies for different neural networks (such as GPT and VGG) via the SBP abstraction, and thus achieves high productivity.
 
 \item We are the first to show that an automated scheme can outperform the manually designed and optimized strategies like Megatron by up to 10\% and 31.1\% on Google Cloud and the Piz Daint supercomputer, respectively.
 
\end{itemize}

Experimental results show that AutoDDL can save the end-to-end training time for Transformer models by up to 31.1\% on GPU clusters equipped with high-performance interconnected networks. The estimated cost to train GPT-3 is more than \$4.6m. This implies that AutoDDL can save about \$1.4m when training such a large model.

\section{Background and Related Work}
\label{sec:bg}
Data parallelism~\cite{parameter_server, distdl, horovod, li2020taming} is a widely-used distributed deep learning approach that partitions the training data across multiple nodes and aggregates gradients during training. Operator parallelism~\cite{wierd_trick, distdl, megatron, spatial, channel} splits specific operators across nodes and communicates intermediate results to satisfy data dependency. Megatron~\cite{megatron} combines both data and operator parallelism, which is designed primarily for Transformer models. OneFlow~\cite{oneflow}, a deep learning framework, proposes SBP abstraction that allows for convenient expression of data, operator, and hybrid parallelisms. AutoDDL builds the automatic strategy search on top of SBP, and we will detail the SBP abstraction in Section~\ref{sec:sbp_sec}.

Most of the operators (such as fully-connected layers and multi-head attention) in deep neural networks are eventually induced to (batched) matrix multiplication operations. Data and/or operator parallelisms on these operators are essentially corresponding to different implementation schemes of Distributed Matrix Multiplication (DMM). In the literature of high-performance computing, distributed matrix multiplication has been intensively studied to reduce the communication cost. Cannon~\cite{cannon1969cellular} and Summa~\cite{summa} are two well-known 2D parallel algorithms, both of which partition the two dimensions of the output matrix and incur $\mathcal{O}\left(\frac{n^2}{p^{1/2}} \right)$ communication cost and $\mathcal{O}\left(\frac{n^2}{p} \right)$ memory cost. These 2D algorithms have been proven to be optimal for the communication cost under the memory budget of $\mathcal{O}\left(\frac{n^2}{p} \right)$~\cite{irony2004communication}. Compared with Cannon that only supports square topology for processes, Summa supports rectangular process topology which is more flexible and general. Summa has been applied in Optimus~\cite{optimus} and Colossal-AI~\cite{ColossalAI} to reduce communication cost for distributed deep learning in memory-constrained scenarios.

On the other hand, the so-called 3D parallel algorithms~\cite{agarwal1995three,aggarwal1990communication,Carma,Cosma} for DMM partition all the three dimensions of the computation space of the matrix multiplication and map the computation to the processes in a cube topology (i.e., $p^{1/3}\times p^{1/3}\times p^{1/3}$). Compared with 2D algorithms, 3D algorithms reduce the communication cost to $\mathcal{O}\left(\frac{n^2}{p^{2/3}} \right)$ but at a cost of $\mathcal{O}\left(\frac{n^2}{p^{2/3}} \right)$ memory consumption. These 3D algorithms are also proven to be optimal in terms of the communication cost~\cite{irony2004communication}. 2.5D parallel algorithms~\cite{solomonik2011communication,Carma,Cosma} are another type of algorithm between 2D and 3D. 2.5D algorithms map the 3D computation space of matrix multiplication to processes in a rectangular cuboid topology (i.e., $p = c\times q\times q$), and incur $\mathcal{O}\left(\frac{n^2}{cq} \right)$ communication cost and $\mathcal{O}\left(\frac{n^2}{q^2} \right)$ memory cost. Note that the communication and memory cost of 2.5D algorithms also lies in between 2D and 3D algorithms. Tuning the parameter $c$ of 2.5D algorithms enables a trade-off between communication cost and memory cost of DMM. 

Although the aforementioned parallel algorithms (2D/3D/2.5D) for DMM have achieved good performance for a single operator~\cite{optimus,wang2022tesseract,ColossalAI}, deep neural network (DNN) training involves both the forward and backward passes, which include data redistribution between different layers. AutoDDL takes into account communications on both the forward and backward paths and properly handles data redistributions between layers, making it a more comprehensive solution for distributed DNN training.


Orthogonal to data and operator parallelisms, pipeline parallelism~\cite{pipedream, pipemare, gpip, li2021chimera, osawa2023pipefisher, dapple, terapip, 2bw, GEMS} is another important approach for distributed deep learning which exploits inter-layer parallelism. The neural network is partitioned in a layer-wise way and consecutive layers form a pipeline stage. Intermediate results are transferred between stages to progress the pipeline. According to the consistency between the weights version and gradients version, pipeline parallelism can be classified into two categories. One is synchronous pipeline parallelism, such as GPipe~\cite{gpip} and DAPPLE~\cite{dapple}. Although synchronous pipeline parallelism is friendly to model convergence, it introduces pipeline bubbles. Chimera~\cite{li2021chimera} is a synchronous approach and alleviates the bubble issue by combining bidirectional pipelines. The other one is asynchronous pipeline parallelism, such as Pipedream~\cite{pipedream} and Pipedream-2BW~\cite{2bw}, which solves the bubble issue but introduces staleness into the training process. Overall, since pipeline parallelism has low communication cost and is relatively easy to implement, it is widely used in distributed training for big models. Deep learning frameworks like FlexFlow~\cite{flexflow} and Tofu~\cite{tofu} did not consider pipeline parallelism, whereas Alpa~\cite{alpa} optimizes pipeline parallelism separately after optimizing data and operator parallelism. Similar to Alpa, AutoDDL optimizes data and operator parallelism based on the SBP abstraction, and then optimizes pipeline parallelism for best performance.

To relieve the burden of machine learning practitioners, automatic deep learning frameworks have also been studied in the literature. Jia et al.~\cite{ExploringHD} have discussed extensively the benefits of exploring different parallel dimensions for distributed deep learning. However, they did not include the partial sum state for exploring parallelism further. Their later work FlexFlow~\cite{flexflow} employs MCMC~\cite{mcmc} to discover a high-performance parallelization and task placement strategy for each operator in a DNN. A recent extension of FlexFlow optimizes tensor fusion and automatic parallelization together~\cite{Unity}. This joint optimization approach is orthogonal and complementary to our strategy of introducing a new search space. It offers a potential future work where the extended AutoDDL search space and tensor fusion could be integrated. Tofu~\cite{tofu} proposes to automatically parallelize operators using the "partition-n-reduction" scheme. This scheme assumes that the operator output on every node can be concatenated or reduced to retrieve the actual result. Pase~\cite{Pase} employs graph algorithms to automate the process of distributed DNN training. Alpa~\cite{alpa} searches a competent parallelization strategy using compilation techniques. However, their approach does not include asymptotically communication optimal algorithms and demonstrates no performance improvements over the expert-optimized scheme (i.e., Megatron) for large-scale experiments. Colossal-AI~\cite{ColossalAI} is another deep learning framework that integrates various DMM algorithms (2D/2.5D/3D) to accelerate DNN training. However, Colossal-AI applies the same process topology for DMM to all layers of a neural network, which lacks the flexibility and misses the opportunities to discover more competent paralellization strategies where different layers have different process topologies. The Colossal-AI team further proposes Colossal-Auto~\cite{ColossalAuto} for automating distributed deep learning with activation checkpointing~\cite{gradients_checkpointing}. However, all aforementioned works are not shown to include the strategies with asymptotically optimal communication cost. Furthermore, AutoDDL enables gradient accumulation to save memory. Unlike activation checkpointing~\cite{gradients_checkpointing} which brings significant recomputation overhead, experimental results show that gradient accumulation has only a small impact on performance but can save a significant amount of memory.


In Table~\ref{comparison}, we compare AutoDDL with existing approaches (manually-optimized schemes and DL frameworks) from four aspects: supported models, communication cost, the cost of parallelization strategy search, and usability. Summa and Megatron are designed to optimize specific operators and models, whereas other approaches such as data parallelism, Colossal-AI, and automatic parallel DNN frameworks (Flexflow, Tofu, Alpa, and AutoDDL) offer broader support for all existing models. Note that Tofu only supports "partition-n-reduction" parallel patterns. AutoDDL, Colossal-AI, and Summa achieve asymptotically optimal communication for DMM, while other approaches do not. However, Colossal-AI and Summa directly implement DMM algorithms for all layers of a DNN, making them less flexible in adapting different strategies to different DNN layers to further improve the performance. In contrast, AutoDDL is the only work that explores specific strategy to each layer, enabling discovering strategies with near-optimal communication cost. In terms of the cost of strategy search for automatic deep learning frameworks, the strategy searching process of FlexFlow and Alpa relies on performance simulation on real GPUs. In contrast, AutoDDL utilizes an accurate performance model combined with low-overhead heuristic algorithms, and the searching process can be finished on CPUs. For complex models and vast search spaces, the performance model-based searching method of AutoDDL is much faster and saves compute resources (i.e., no requirement for GPUs). In terms of usability, data parallelism and automatic frameworks are easier to use. Colossal-AI is user-friendly with some restrictions, allowing only 2D or 3D operator parallelisms with squared or cubic numbers of nodes. Summa and Megatron require users to determine the 2D process topology and only support specific models. 


There are other techniques for reducing communication overhead that are orthogonal to AutoDDL. For example, gradient sparsification~\cite{topk0, topk1, topk2} only communicates the largest k gradients to reduce the communication volume, while gradient quantization~\cite{quant0, quant1, quant3} reduces the number of bits for the gradient values by using lower precision. These communication-reducing techniques can be integrated into AutoDDL by factoring in the reduced communication volume when evaluating the performance of different parallelization strategies.  
\section{Communication Performance Model}
\label{sec:comm}
Collective communications play a crucial role in distributed deep neural network training. The most widely used collective communication operation is AllReduce~\cite{allreduce}, which is essential for synchronizing weight gradients in data parallelism. In addition to AllReduce, AutoDDL also requires other collective operations for operator parallelism, such as AllGather, ReduceScatter, and AlltoAll. 


\subsection{Collective Performance based on the Alpha-Beta Model}
The alpha-beta model is a well-known model to simulate communication time. In this model, $\alpha$ represents the network latency, and $\beta$ represents the inverse network bandwidth. Communicating an n-byte message takes $\alpha + n\beta$ time. Some collective operations, such as ReduceScatter and AllReduce, also involve computation, but we omit the computation cost in the model as it is negligible compared with the communication cost. Studies conducted by MPICH~\cite{thakur2005optimization} have employed the alpha-beta model to analyze the performance of collective communication based on message sizes. In our performance model, we utilize the complexity of bandwidth-optimal, long-message-friendly collective algorithms, as large model training often involves the transmission of lengthy messages. The algorithms employed encompass ring algorithms for AllReduce and AllGather, along with pairwise exchange algorithms for ReduceScatter and AlltoAll, as summarized in Table~\ref{table:collectives}.

\begin{table}
  \centering
  \renewcommand{\arraystretch}{1.6}
  \begin{tabular}{lll}
    \toprule

    Collective &  Communication Complexity \\
    \hline 
    AllReduce & \((2p-1)  \alpha +  2 \frac{p-1}{p} n\beta \) \vspace{1pt}\\ 
    \hline
    AllGather/ReduceScatter &  \( (p-1)  \alpha +  \frac{p-1}{p}  n\beta\) \vspace{1pt}\\ 
    \hline
    AlltoAll &  \( (p-1)  \alpha +  n\beta\)
    \vspace{1pt}\\
 
    \bottomrule
  \end{tabular}
  \vspace{-5pt}
  \caption{Communication Complexity of Bandwidth-Optimal MPI Collectives. N is the message size for ReduceScatter's inputs and other collectives' outputs. \vspace{-8pt}}
  \label{table:collectives}
\end{table}

\subsection{Inferring Network Parameters on Real World Systems}
\label{sec:infer_network}
Instead of relying on the peak latency and bandwidth values reported by hardware manufacturers, we develop a more practical alpha-beta model for collective communication by estimating the $\alpha$ and $\beta$ parameters in a real-world environment. To this end, we construct communication models for two different platforms: the Piz Daint Supercomputer, which is a Multi-Node-Single-GPU system, and a Google Cloud GPU platform (two DGX A100 servers and each server has 16 A100 GPUs), which is a Multi-Node-Multi-GPU system. We describe the configurations of these platforms in Section~\ref{sec:exp_sec}.

We then conduct experiments to measure the actual runtimes of four collective operations (AllReduce, AllGather, AlltoAll, and ReduceScatter) with various message sizes and numbers of processes. We repeat each experiment 1000 times to obtain reliable statistics. Based on the data collected, we can formulate a linear system in terms of $\alpha$ and $\beta$ for each network, as shown in Table~\ref{table:collectives}. We use the least squares method to estimate the values of $\alpha$ and $\beta$ from the collected data.

\subsection{Ranking of different parallel strategies via the communication model}

The purpose of the communication model is to rank various parallel strategies, facilitating the selection of a strategy with low communication costs. In AutoDDL, each DNN layer has parallelization strategies defined by SBP notation, as detailed in Section~\ref{sec:sbp_sec_strategies}. The diverse strategies involve data redistributions between layers and gradients synchronizations, with predetermined information on all collective communication types, the number of GPUs involved, and message sizes, as outlined in Section~\ref{sec:sbp_sec_cost}. This information enables the prediction and ranking of different strategies through the application of the alpha-beta model.

\begin{figure}[t]
\centering
    \includegraphics[width=0.45\textwidth]{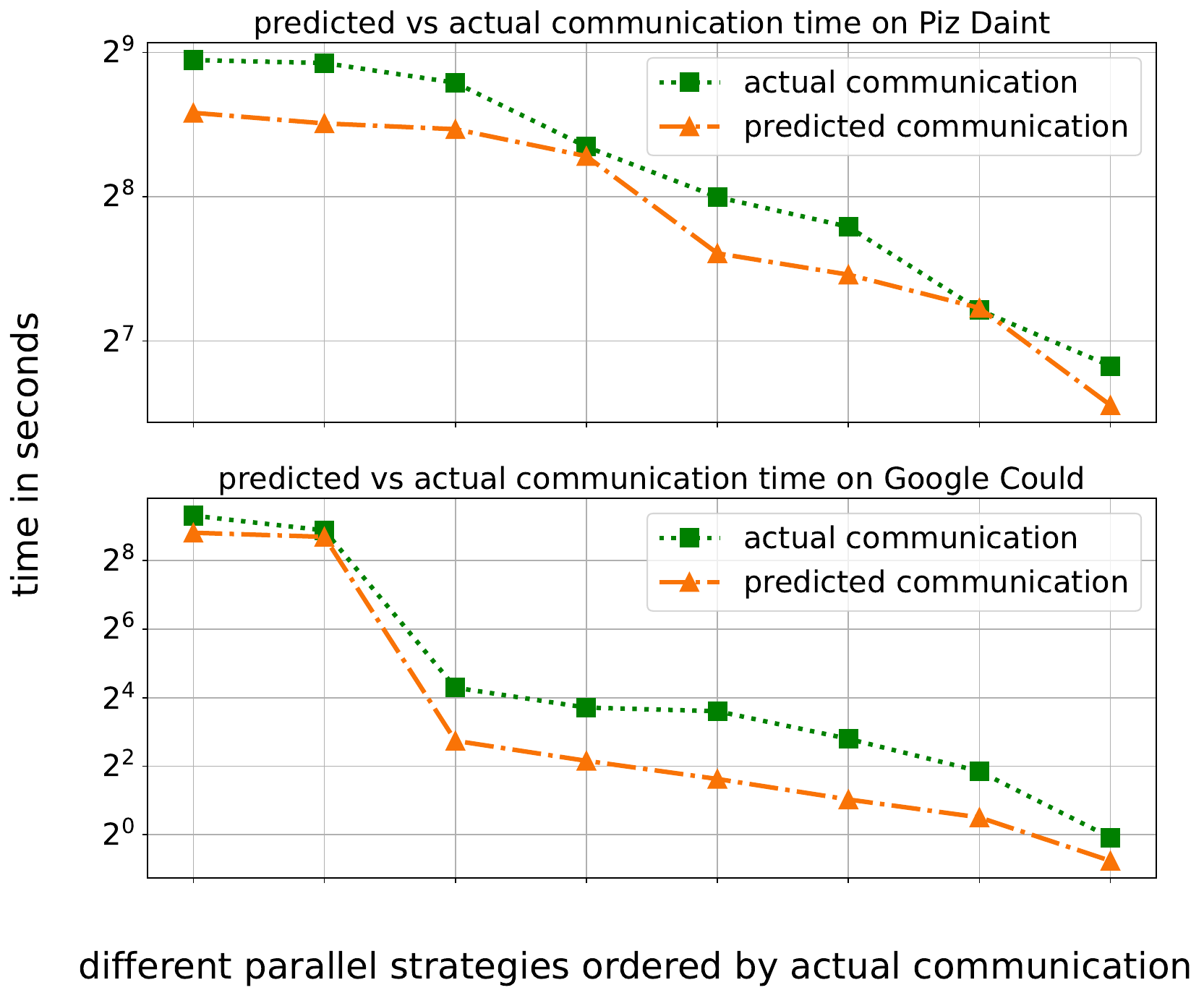}
    \vspace{-5pt}
\caption{Actual communication time versus the performance model's prediction for eight different parallelization strategies of a Transformer model. \label{fig:performance_model} \vspace{-8pt}}
\end{figure}

In Figure~\ref{fig:performance_model}, we show the actual communication costs versus the performance model's prediction for a Transformer model on Piz Daint and the Google Cloud GPU platform. The experiments are conducted on 32 single-GPU nodes on Piz Daint and 2 eight-GPU nodes on Google Cloud. The Google Cloud GPU platform represents a heterogeneous Multi-node-Multi-GPU environment. In this case, different $\alpha$ and $\beta$ parameters are used for inter- and intra-node communications, respectively, and the details of modeling strategies in such an environment are outlined in Section~\ref{sec:sbp_sec_strategies}. The parallelization strategies illustrated in Figure~\ref{fig:performance_model} are presented in descending order based on the measured communication costs. It is evident that the performance model accurately captures the ranking of communication time on both platforms.

\vspacegraph
\section{The AutoDDL Strategy Search}
\label{sec:sbp_sec}
The success of AutoDDL heavily relies on the automatic selection of parallelization strategies for distributed deep learning. To achieve this, we first define the AutoDDL search space, which is based on OneFlow's native SBP abstraction and interface. In Section~\ref{sec:sbp_sec_oneflow}, we explain the SBP abstraction, and in Section~\ref{sec:sbp_sec_strategies}, we describe the AutoDDL's search space. In Section~\ref{sec:sbp_sec_cost}, we demonstrate that the search space of AutoDDL enables exploring strategies with near-optimal communication cost. In Section~\ref{sec:sbp_sec_search} and Section~\ref{sec:sbp_sec_strategy}, we present the heuristic search algorithm used in AutoDDL to efficiently discover near-optimal strategies for end-to-end DNN models.

\vspacegraph
\begin{figure}[t]
\centering
    \includegraphics[width=0.49\textwidth]{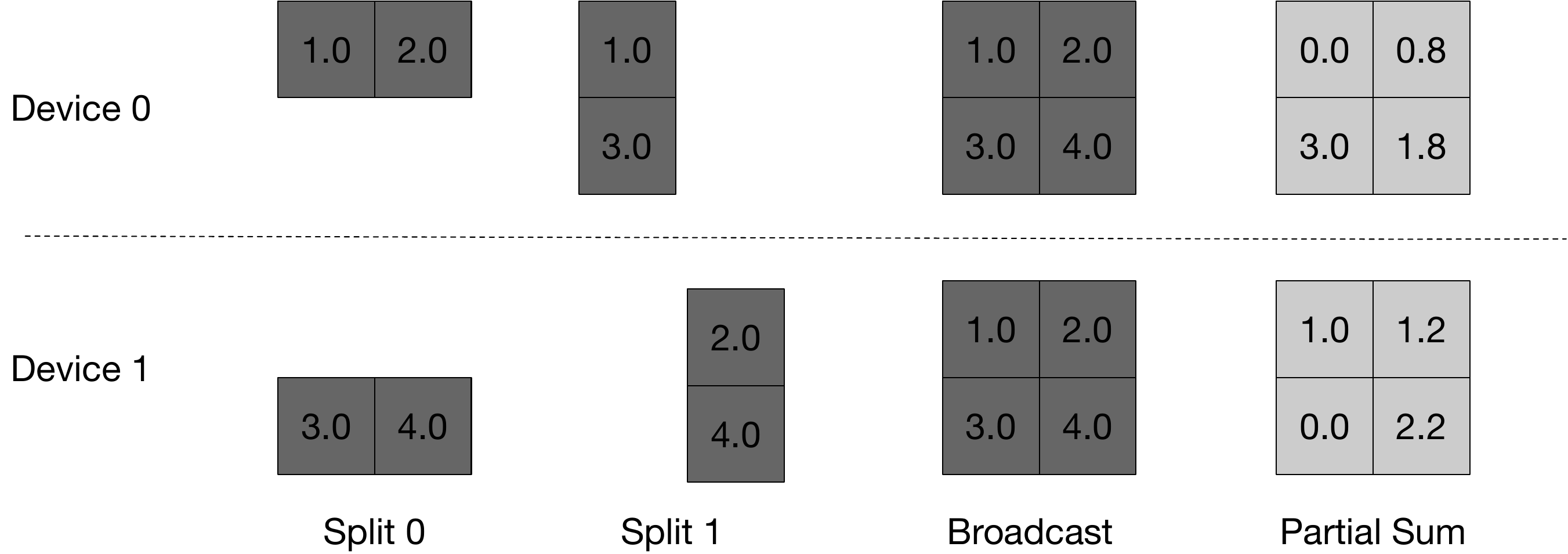}
    \vspace{-0pt}
\caption{Basic SBP of a 2d tensor on a group of two devices. Bright color represents a partial sum: the elements must be summed to retrieve the original tensor. \label{fig:sbp_basic} \vspace{-8pt}}
\end{figure}

\subsection{OneFlow's native support for SBP}
\label{sec:sbp_sec_oneflow}
OneFlow provides an abstraction called SBP that describes the states of distributed tensors. With SBP, an n-dimensional tensor can be Split, Broadcasted, or Partial-sum along all n dimensions on different devices (see Figure \ref{fig:sbp_basic}).
\begin{lstlisting}[language=Python,
    keywordstyle=\color{blue}\ttfamily,
    stringstyle=\color{red}\ttfamily,
    breaklines=true, caption=OneFlow SBP Initialization Example, label=code_init,
    basicstyle=\ttfamily\footnotesize, label=code_init]
#initialising a 2d tensor w
#w has distribution (S1:2, S0:3, B:4)
#device_id is a list of devices
import oneflow
distribution = ["S(1)", "S(0)", "B"]

with oneflow.scope.placement("gpu", device_id, (2, 3, 4)):
    w = oneflow.get_variable(
            name="2d_tensor",
            shape=(6, 4),
            dtype=oneflow.float,
            trainable=True,
            parallel_distribution=distribution,
    )
\end{lstlisting}
When compared to the GShard~\cite{gshard} abstractions, OneFlow's Broadcast state corresponds to the Replicate state in GShard. Additionally, OneFlow's Split states generalize the Split and Shard states in GShard. Notably, the SBP abstraction in OneFlow includes one more state, Partial-sum (\textit{P}), which extends the search space and potentially enables better parallelization strategies.

The OneFlow framework offers built-in support for distributed tensor initialization and redistribution. With a total of 24 devices available, Listing~\ref{code_init} presents the code for initializing a 2D tensor. The 24 devices are arranged in a 3D mesh topology, e.g., $2 \times 3 \times 4$. The two dimensions of the 2D tensor are split across the first two dimensions of the 3D mesh topology. The entire 2D tensor is then broadcasted across the third dimension of the 3D mesh topology.

Listing~\ref{code_boxing} presents a code example demonstrating tensor redistribution for tensor x. Initially, total 32 devices are arranged in a 2D mesh topology ($4 \times 8 $). Tensor x is in the state of partial-sum along one dimension of the 2D mesh (i.e., 4 devices), with the dimension 1 of tensor x being split across the other dimension of the 2D mesh (i.e., 8 devices). Following the redistribution process (reduce-scatter in the forward path), the dimension 1 of tensor x is now split across all 32 devices. 

\begin{lstlisting}[language=Python,
    keywordstyle=\color{blue}\ttfamily,
    stringstyle=\color{red}\ttfamily,
    breaklines=true, caption=SBP Initialization Example, label=code_init,
    basicstyle=\ttfamily\footnotesize, caption=OneFlow Tensor Redistribution Example, label=code_boxing]
# a tensor redistribution example 
#device_id is a list of devices
import oneflow

with oneflow.scope.placement("gpu", device_id, (8, 4)):
    #previous x has distribution (S1:8, P:4)
    x = oneflow.hierarchical_parallel_cast(x,parallel_distribution=["S(1)", "P"])
    
with oneflow.scope.placement("gpu",device_id):
    x = oneflow.hierarchical_parallel_cast(x,parallel_distribution=["S(1)"])
#now x has distribution (S1:32)
\end{lstlisting}

AutoDDL is able to directly utilize OneFlow's built-in APIs for distributed tensor initialization and tensor redistribution (similar to Listing~\ref{code_init} and Listing~\ref{code_boxing}). Any parallelization strategy selected by AutoDDL can be easily implemented using the OneFlow APIs. This approach removes the need of explicitly handling the tensor placement and the communication for distributed tensors, significantly alleviating the burden on programmers.

\subsection{Parallel Strategies}
\label{sec:sbp_sec_strategies}
The SBP tensor abstraction in OneFlow facilitates the description of the search space of parallelization strategies for distributed training. These strategies can be adapted to different layers, with each operator having a unique distributed tensor state. Certain operations come with additional constraints on input split dimensions. 


\vspace{4pt}\noindent\textbf{General SBP Layer.} With the help of AutoDDL, engineers can effectively implement parallelization strategies by constructing a sequence of computations and SBP tensor redistributions within the search space. AutoDDL then simulates the overhead associated with each SBP redistribution and identifies the optimal strategy that minimizes communication costs. Moreover, performance engineers have the flexibility to extend communication patterns beyond SBP, such as invent their own parallel implementations or incorporating existing patterns like the halo exchange techniques for CNNs~\cite{spatial}. For instance, if there exists a precise performance model for halo exchange, AutoDDL has the capability to recognize halo exchange as an implementation that has its input and output tensors in a SBP state. Subsequently, AutoDDL can seamlessly integrate halo exchange into its search space.

\begin{figure}[t]
    \centering
    \includegraphics[width=0.49\textwidth]{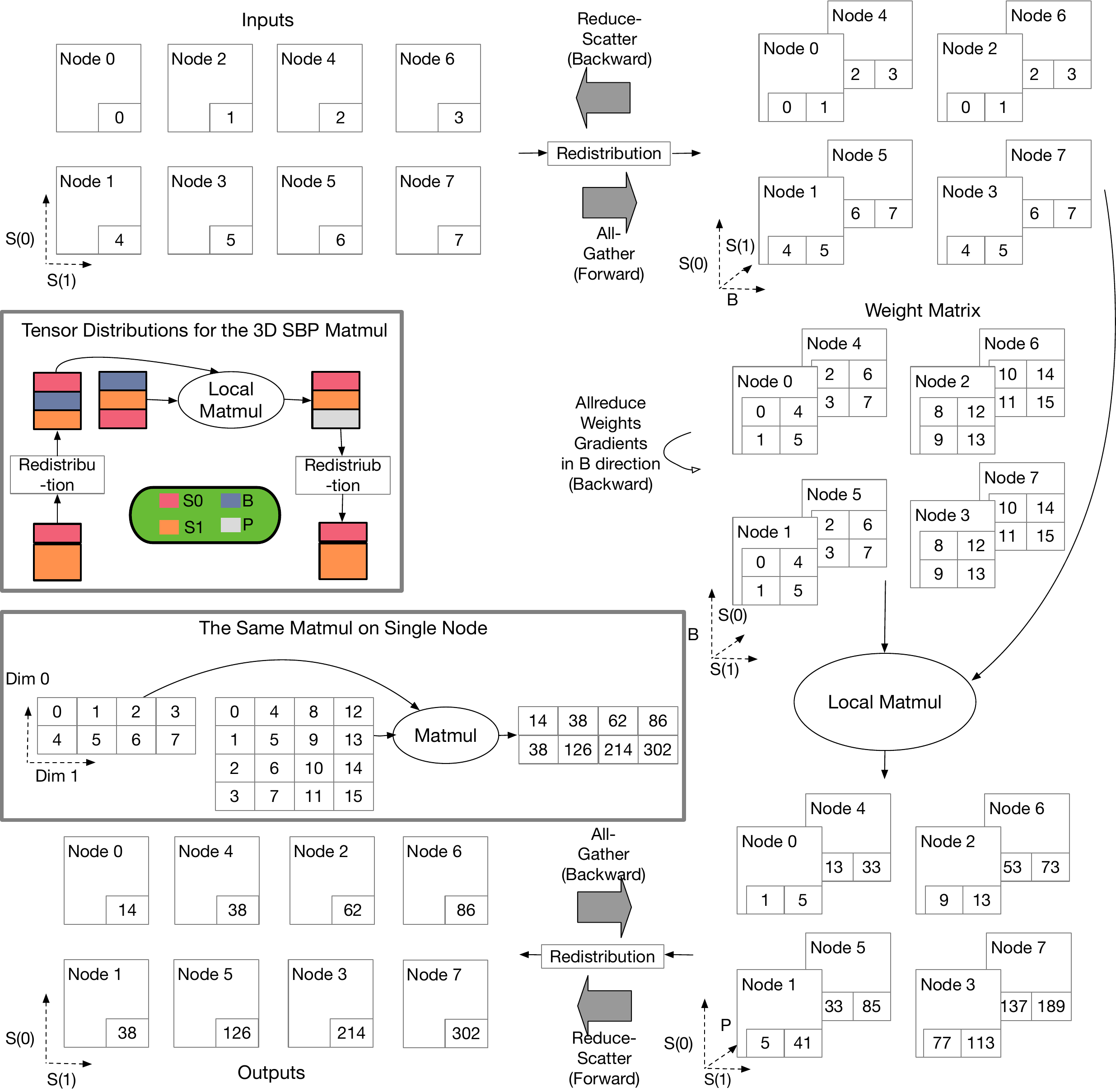}
    \vspace{-0pt}
\caption{shows a 3d SBP DMM example. The tensor distributions are shown in color, and the same matrix multiplication on a single node is included for comparison.
 \label{fig:3d_matmul} \vspace{-8pt}}
\end{figure}

\vspace{4pt}\noindent\textbf{Distributed Matrix Multiplication based on SBP.} Matrix multiplication (Matmul) is a fundamental operation in deep learning. Our SBP matrix multiplication algorithm can be described as follows: The input matrix $I$, weight matrix $W$, and the output matrix $O$ are all two-dimensional tensors. We assume that $p=p_0\times p_1\times p_2$ devices are available. For the matrix multiplication $IW = O$, if the input matrix $I$ (after the first redistribution in Figure~\ref{fig:3d_matmul}) has an SBP distribution of (S0: $p_0$, B: $p_1$, S1: $p_2$) and the initialized weight matrix $W$ has an SBP distribution of (B: $p_0$, S1: $p_1$, S0: $p_2$), then the partial results of the matrix multiplication is performed locally on each device (Local Matmul in Figure~\ref{fig:3d_matmul}). The output matrix $O$ will have an SBP distribution of (S0: $p_0$, S1: $p_1$, P: $p_2$). At last, a reduce-scatter operation is required to obtain the final results of the matrix multiplication. Figure~\ref{fig:3d_matmul} shows an example of the SBP Matmul with actual tensor distributions. Note that if both $p_1$ and $p_2$ are equal to 1, the algorithm reduces to data parallelism.


For an SBP layer, we first specify the input SBP distribution and constrain its output distribution. Our AutoDDL framework can then automatically conduct data redistribution between two consecutive layers. Let us consider an example of the SBP matrix multiplication. For consecutive SBP matrix multiplication operations $M_{i}$ and $M_{i+1}$, it is necessary to ensure that the output of $M_{i}$ has the correct SBP distribution to serve as the input to $M_{i+1}$. If the output of $M_{i}$ is not compatible with the input of $M_{i+1}$, implicit data redistribution is necessary to convert the tensor to the correct SBP distribution as explained in the following. If the output of $M_i$ has an SBP distribution of (S0: $p_0$, S1: $p_1$, P: $p_2$) and $M_{i+1}$ requires an input with an SBP distribution of (S0: $p_0'$, B: $p_1'$, S1: $p_2'$), where $p_0=p_0'$, the required tensor redistribution path is: (S0: $p_0$, S1: $p_1$, P: $p_2$) -> (S0: $p_0$, S1: $p_1*p_2$) -> (S0: $p_0'$, B: $p_1'$, S1: $p_2'$). The "->" symbols denote the appropriate redistributions, which translate to collective communications in the forward/backward path. This example with two data redistributions is illustrated in Figure~\ref{fig:3d_matmul}, which involve ReduceScatter and Allgather operations in both the forward path and the backward path.

For the general case where $p_0\neq p_0'$ or $p_1\neq p_2'$, the required change in data distribution is as follows: (S0: $p_0$, S1: $p_1$, P: $p_2$) -> (S0: $p_0$, S1: $p_1*p_2$) -> (S0: $p_0'$, S1: $p_1'*p_2'$) -> (S0: $p_0'$, B: $p_1'$, S1: $p_2'$). Additionally, AlltoAll collectives are required for the (S0: $p_0$, S1: $p_1*p_2$) -> (S0: $p_0'$, S1: $p_1'*p_2'$) transformation.

\vspace{4pt}\noindent\textbf{Strategies for heterogeneous interconnected networks.}
So far we have explained the AutoDDL parallelization strategies for Multi-Node-Single-GPU machines in previous paragraphs, where the interconnected network is homogeneous. The tensor distributions and strategies in the Multi-Node-Multi-GPU case are similar to those in the Multi-Node-Single-GPU case. However, instead of using numbers to denote SBP, we use tuples to represent the number of nodes and the number of GPUs per node for each SBP dimension to better utilize the heterogeneous interconnected network. For example, the input matrix for an SBP matmul is of the form (S0: $(p_0, p_0’)$, B: $(p_1,p_1’)$, S1: $(p_2,p_2’)$) for a system with $p=p_0\times p_1 \times p_2$ nodes and each node with $p’=p_0’ \times p_1’ \times p_2’$ GPUs per node. For example, in a 2D scenario with 2 nodes, each node equipped with 4 GPUs, the configuration (S0:(1, 2), B:(2, 2)) implies that the initial split (along the 0th dimension) of the data is replicated on GPUs {(0, 0), (0, 2), (1, 0), (1, 2)}, while the second split of the data is replicated on GPUs {(0, 1), (0, 3), (1, 1), (1, 3)}.

If a communication only involves intra-node GPUs (the first entry in the corresponding tuple is 1), the performance model is instantiated by the bandwidth and the latency of intra-node NVlink. Otherwise, the communication is deemed to be on the slower inter-node network. Taking data redistribution on two nodes and each node with 4 GPUs as an example: in the case of (S0:(1, 2), B(2, 2)) -> B:(2, 4), since the tensor is split on two GPUs in the same node, only intra-node communication is required to gather the whole tensor; on the other hand, in the case of (S0:(2, 1), B(1, 4)) -> B:(2, 4), since the tensor is split on two different nodes, inter-node communication is required to gather the whole tensor.

\subsection{Communication Cost Analysis} 
\label{sec:sbp_sec_cost}
Once the different layers with varying parallel strategies have been defined and combined, the communication for the entire deep neural network (DNN) can be separated into a sequence of collectives. Each collective operation can then be analyzed in accordance with Section~\ref{sec:comm}.

\vspace{4pt}\noindent\textbf{Sources of Communication.}
Communication within the AutoDDL search space occurs from two sources. The first type of communication involves data redistribution, as discussed in Section~\ref{sec:sbp_sec_strategies}. The second type of communication involves weight gradient synchronization, where any replicated (Broadcast) weight matrix W receives only a portion of the gradients during the backward pass. The associated weight gradients are implicit of Partial-Sum state and an AllReduce operation to accumulate the gradient information is automatically performed by OneFlow.

\vspace{4pt}\noindent\textbf{Asymptotic Communication for SBP Matmul.}
It is a well-known fact that the communication lower bound for an n*n matrix with p devices, each consuming $\Theta(\frac{n^2}{p^{2/3}})$ memory per device, is $\mathcal{O}(\frac{n^2}{p^{2/3}})$~\cite{lower_bound}. Figure~\ref{fig:3d_matmul} provides an example of a complete 3D SBP matrix multiplication. One can observe from the figure that the data volume is divided by p and then increased by the dimension of Partial Sum (P) or Broadcast (B), which is less than $\max(p_1, p_2, p_3)$, before or after redistributions. And the gradient synchronizations and data redistributions have communication volumes no larger than the maximal data volume of the local tensor, before or after redistributions. To derive an upper bound for the data volumes, we can select $p_1$, $p_2$, and $p_3$ (all two in the example figure) to be in $\mathcal{O}(p^{1/3})$, the communication volume and memory occupation will both be in $\Theta(\frac{n^2}{p^{2/3}})$ with the tensors of size $\mathcal{O}(n^2)$. This example demonstrates that our search space includes asymptotically optimal 3D distributed matrix multiplication.

\subsection{Heuristic Search Algorithms}
\label{sec:sbp_sec_search}
The AutoDDL search space can potentially include all possible SBP parallelization strategies for each layer, leading to a large number of potential solutions. To reduce the search space, we leverage the GPU memory limit. It is important to note that different strategies can have different implications for memory consumption, and some strategies may exceed the GPU memory capacity and cause out-of-memory (OOM) errors. To avoid this issue, we utilized the OneFlow framework to figure out the exact memory usage on CPU before training on GPUs. We exclude any strategies that exceed the GPU memory limit to ensure successful execution during training.

Despite the search space can be reduced to some extend by utilizing the memory capacity constraint, the search space can still be enormous for complex models trained on thousands of large-scale nodes. An exhaustive search would require iterating over the entire search space, which can be time-consuming. In this subsection, we investigate two heuristic algorithms, including Metropolis-Hastings and coordinate descent, and compare them with exhaustive search and random search.


\vspace{4pt}\noindent\textbf{The Metropolis-Hastings Algorithm.} 
We define the sampling probability for each strategy as follows:
\[
P(S) \propto exp(-\beta \cdot Cost(S))
\]
where $\beta$ is a constant chosen by the user, and $\text{Cost}(S)$ is the cost associated with strategy $S$. We use the Metropolis-Hastings algorithm~\cite{Hasting} to sample from the space of possible strategies. Starting from an initial strategy, the algorithm proposes a new strategy by randomly changing the SBP distribution of one layer. The proposed strategy is accepted with probability
\[
min(1, P(S_{new}) / P(S_{old}))
\]
where $S_\text{old}$ and $S_\text{new}$ are the current and the proposed strategies, respectively. The algorithm repeats this process indefinitely, creating a chain of strategies.

The Metropolis-Hastings algorithm, which is utilized in Flexflow's MCMC simulation~\cite{flexflow}, employs an acceptance rule that is based on the difference in cost between the current and proposed strategies. Specifically, if a new strategy has a lower cost than the current one, it is immediately accepted. However, if a new strategy has a higher cost, its acceptance probability decreases exponentially with the difference in cost to the old strategy. This allows the algorithm to occasionally accept higher-cost strategies, which can help avoid getting stuck in local minima. However, in most cases, Metropolis-Hastings will accept a strategy that is at least as good as the current one.


\vspace{4pt}\noindent\textbf{The Coordinate Descent Algorithm.}
The coordinate descent algorithm is a method for optimizing the search space by iteratively updating the SBP distributions for each layer. It starts with an initial strategy and optimizes parallelization strategy of each layer by minimizing the total cost, with the other layers' distributions fixed: a strategy S for a neural network with L layers can be written as $S = [s_j]_{j=0}^L$ with $s_j$ the strategy for layer j. One update step of coordinate descent is that for every layer i
\[
S[i] = \argmin_{s_i} Cost([s_j]_{j=0}^L) 
\]

This algorithm efficiently explores a large discrete search space, as demonstrated in previous work~\cite{coord_discrete}. It guarantees monotonic cost reduction through greedy line searches along a single layer's strategy space. In our implementation, when a fixed point is reached—meaning the cost no longer decreases with the current initial guess—the algorithm records the best strategy and begins anew with a different initial guess. This approach enables the algorithm to avoid local minima by exploring various regions of the search space.

\vspacegraph
\begin{figure}
\centering
    \includegraphics[width=0.48\textwidth]{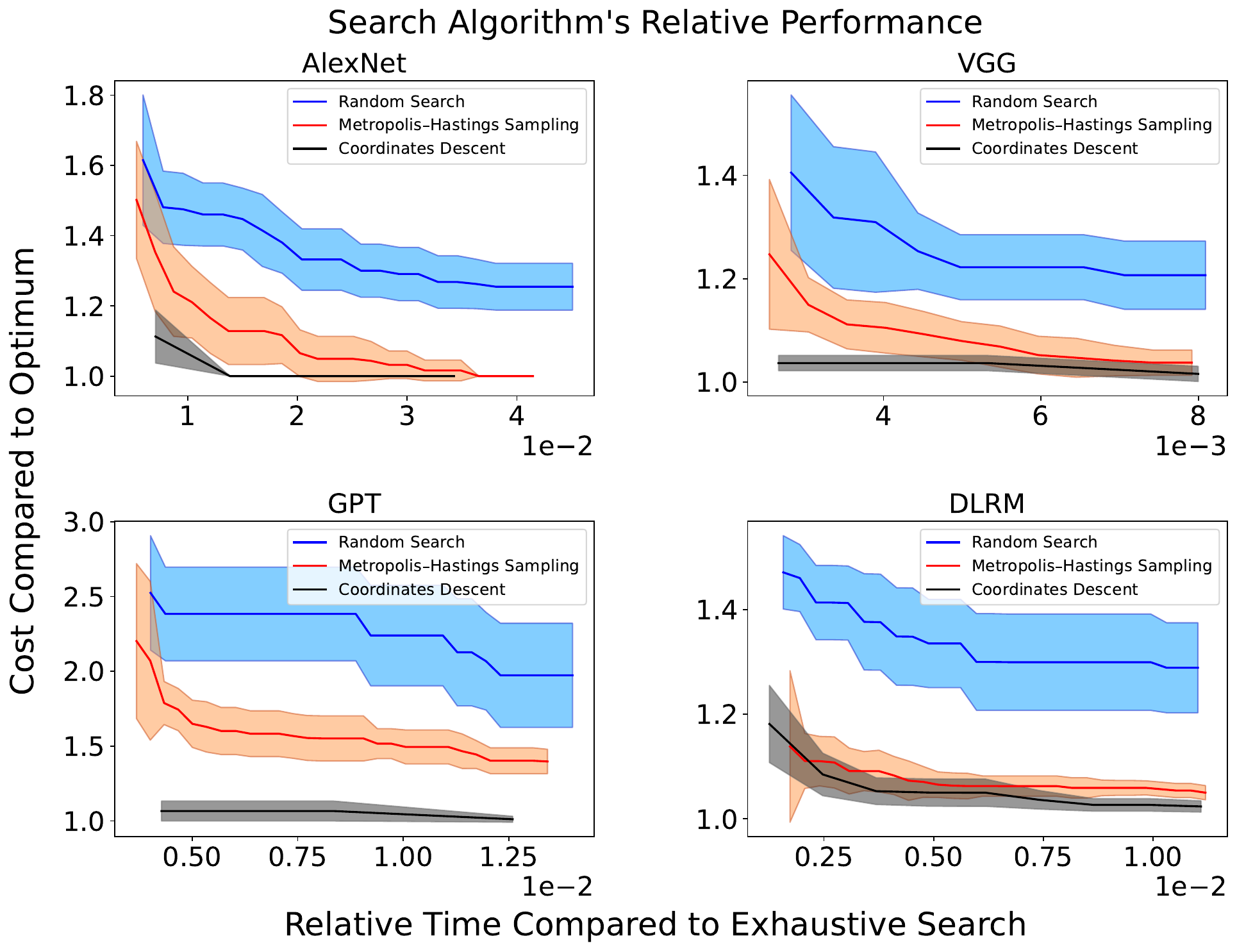}
    \vspace{-5pt}
\caption{Performance comparison of random search, exhaustive search, metropolis-hastings, and coordinate descent algorithms. The shaded area shows the 90\% confidence interval over ten runs.\label{fig:search_algo} \vspace{-8pt}}
\end{figure} 

\vspace{4pt}\noindent\textbf{Evaluation.}
We compare Metropolis-Hastings and coordinate descent algorithms with exhaustive and random search methods for selecting strategies across four neural network types: AlexNet~\cite{krizhevsky2012imagenet}, VGG~\cite{vgg}, GPT~\cite{brown2020language}, and DLRM~\cite{naumov2019deep}.
 
The experiments were conducted on Piz Daint Supercomputer's Intel® Xeon® E5-2695 v4 @ 2.10GHz CPU. Figure \ref{fig:search_algo} presents the relative runtime of each algorithm compared to an exhaustive search, where a value of 1 on the y-axis indicates the optimal solution obtained by an exhaustive search. The random search algorithm cannot find a close-to-optimal strategy within the limited runtime budget. The Metropolis-Hastings algorithm, previously utilized in Flexflow, ranks as the second-best performer. Surpassing all other methods, the coordinate descent algorithm quickly identifies near-optimal strategies for various models. Compared to exhaustive search, it achieves this in significantly less time—two orders of magnitude less—while maintaining a gap of less than 3\% from the optimal strategy. Our experiments showed that a single run of coordinate descent converges quickly, allowing the algorithm to perform hundreds of runs with different initial points in a short time. This rapid iteration is crucial for escaping local minima.

AutoDDL uses the coordinate descent algorithm by default due to its superior empirical performance. Thanks to efficient heuristic algorithms, our AutoDDL strategy search efficiently scales to more complex models across multiple compute nodes at a low cost.

\subsection{Searching An End-to-End Strategy}
\label{sec:sbp_sec_strategy}
\begin{figure}
\centering
    \includegraphics[width=0.48\textwidth]{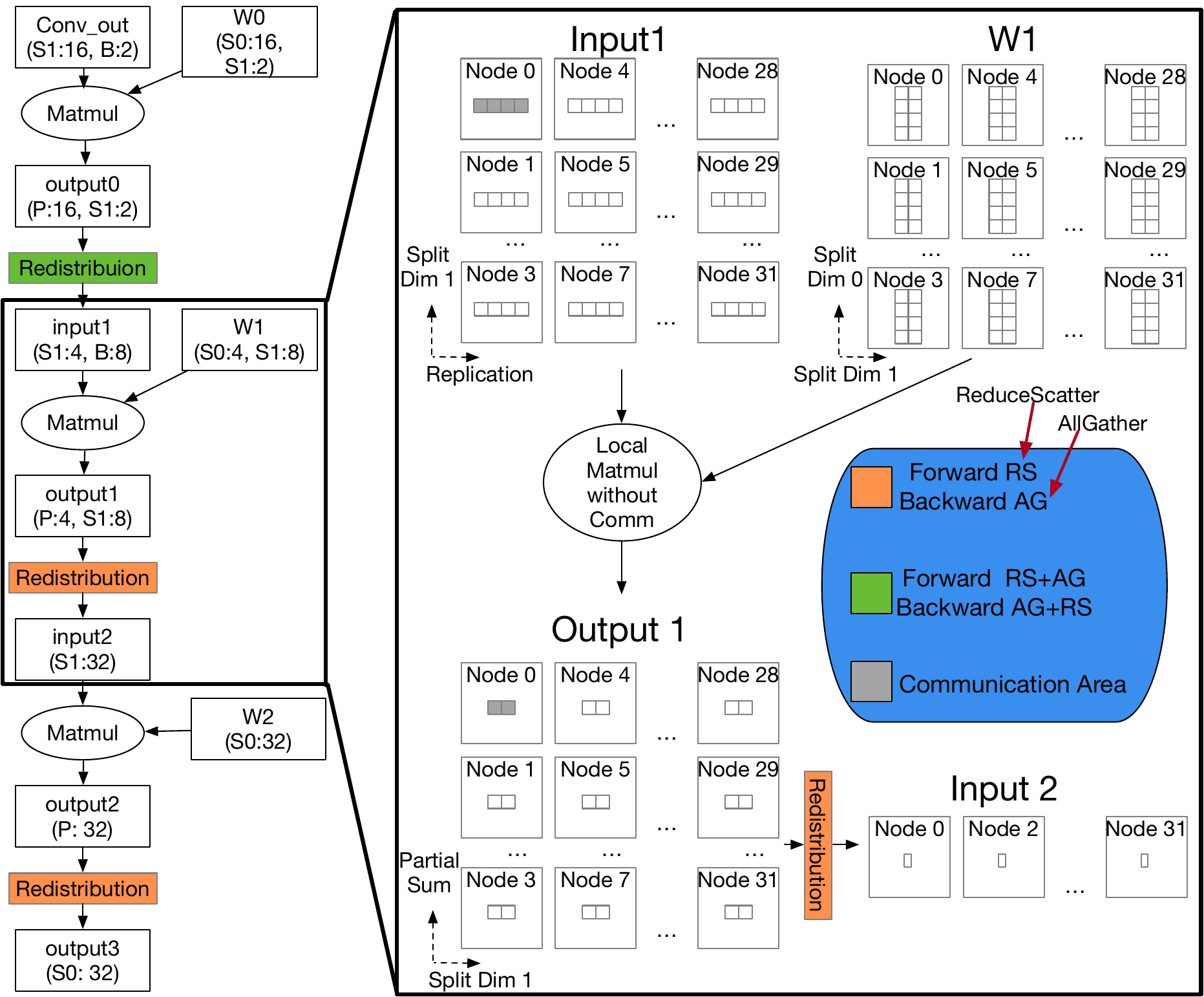}
    \vspace{-0pt}
\caption{AutoDDL for the last three dense layers of vgg13 on 32 GPUs. The second matrix multiplication and the tensor redistributions are shown in detail. \label{fig:strategy_example} \vspace{-8pt}}
\end{figure}

With the search space being defined and searching process being accelerated by the heuristic algorithms, one can efficiently obtain a near-optimal parallelization strategy for end-to-end neural networks. Taking the VGG model used in our experiments as an example, AutoDDL chooses to only use data-parallelism for all the convolution layers of VGG, with (S0: p) for input and output tensors. For the last three dense layers of VGG, Figure~\ref{fig:strategy_example} presents the strategy selected by AutoDDL. In the selected strategy, the weight matrices W1, W2, and W3 are not replicated (no broadcast state), so that zero weight gradient synchronization overhead occurs. This is reasonable because the model dimensions are significantly larger than the batch size for VGG, and avoiding weight gradient synchronization helps to reduce the overall communication overhead. The three dense layers adapt their strategies to different model dimensions that they have. One 2d model parallel matrix multiplication is shown in detail on the right side of Figure~\ref{fig:strategy_example}. In contrast to the 1d model parallelism used by Megatron, Flexflow or Alpa, the 2d model parallel strategy can reduce communication with increased number of nodes, while completely eliminating weight synchronization overheads of data parallelism.

While Figure~\ref{fig:strategy_example} demonstrates the optimal parallelization strategy for VGG involving only two dimensions, it is worth noting that for other layers like the attention layer in Figure~\ref{fig:transformer_example}, 3D parallelization strategy is preferred.

\vspacegraph

\section{Experiments}
\label{sec:exp_sec}
We conduct the evaluation on two deep neural network architectures (Transformers and VGG) using both Multi-Node Single-GPU and Multi-Node Multi-GPU systems. We compare the performance of AutoDDL against the baselines including data parallelism, and highly-optimized Megatron and Summa algorithms. Here Summa algorithm is from Optimus~\cite{optimus}. Since both Megatron and Summa are 2D algorithms where the 2D process topology has a big impact on the performance, both baselines are optimized by selecting the best 2D topology given the total number of processes. For the Multi-Node Multi-GPU system, intermediate results are communicated within a compute node (i.e., among GPUs connected by NVLink which has high bandwidth and low latency), while model updates are transferred over the interconnected network across nodes. We exclude Colossal-AI from the baselines since it imposes a requirement of using a cubic number of processes, despite its potential to achieve asymptotically optimal communication cost using its 3D algorithm. The restrict requirement for the number of processes makes it unsuitable for the majority of our experiments. Additionally, Alpa has not demonstrated performance advantages over the expert-optimized schemes (such as Megatron). Therefore, we use the manually-optimized Megatron as a stronger baseline and thus excludes Alpa from the baselines. All experiments are repeated ten times, and the error bars represent the 99\% confidence interval around the mean, as recommended by the benchmarking standards~\cite{benchmarkSci}.

\subsection{Neural Networks for Evaluation}

\vspace{4pt}\noindent\textbf{The VGG Model.}
We conduct experiments using VGG13~\cite{vgg} and increase the batch size with the number of GPUs. Memory capacity constraint is not a limiting factor for most convolutional neural networks (CNNs), including VGG13, so no gradient accumulation is applied for VGG13.

The convolution layers in VGG13 contribute most of the computation but only a small portion of the weights, making data parallelism suitable for them. On the other hand, the last three dense layers are responsible for most of the communication, and we apply the best parallelization strategy discovered by AutoDDL for these layers. After the outputs of the convolution layers have been transformed to the appropriate distributions, a Megatron-like 2d algorithm (where the reduction dimension is split) can be applied to the last three dense layers.

Although the Summa algorithm helps to alleviate memory redundancy issues, it does not provide the best performance for non-memory-constrained cases like CNNs. Our empirical evaluations show that Summa is an order of magnitude slower than other methods on VGG. Therefore, we omit Summa from the plots of experimental results on VGG.

\vspace{4pt}\noindent\textbf{The Transformer Model.}
The Transformer model~\cite{vaswani2017attention} has several parameters that determine the model size, including the model dimension, the number of attention heads, and the attention head dimension. In our experiments, we use the model sizes from the Megatron-lm paper~\cite{megatron}.

Typically, the mini-batch for language model pre-training involves millions of tokens, which lead to a lot of intermediate results and put pressure on memory. To overcome this, we use gradient accumulation, which divides each mini-batch into micro-batches and accumulates their gradients. To achieve high computation efficiency, we use the largest possible micro-batch.

\subsection{Results on a Multi-Node Single-GPU Machine}
\vspace{4pt}\noindent\textbf{System Configurations.}
The experiments are conducted on the Piz Daint supercomputer. On Piz Daint, each Cray XC50 compute node contains an Intel Xeon E5-2690 CPU and one NVIDIA P100 GPU with 16 GB global memory. The compute nodes of Piz Daint are connected by a Cray Aries interconnect network in a Dragonfly topology. We perform the training in single precision, since the P100 GPUs do not have tensor cores which limits the benefits of mixed-precision. To handle communication, we use the default NCCL library for OneFlow. To obtain our AutoDDL strategy, we build communication performance models for Piz Daint as described in Section~\ref{sec:comm} and apply the strategy as outlined in Section~\ref{sec:sbp_sec}. 

\vspacegraph
\begin{figure}[t]
\centering

    \includegraphics[width=0.48\textwidth]{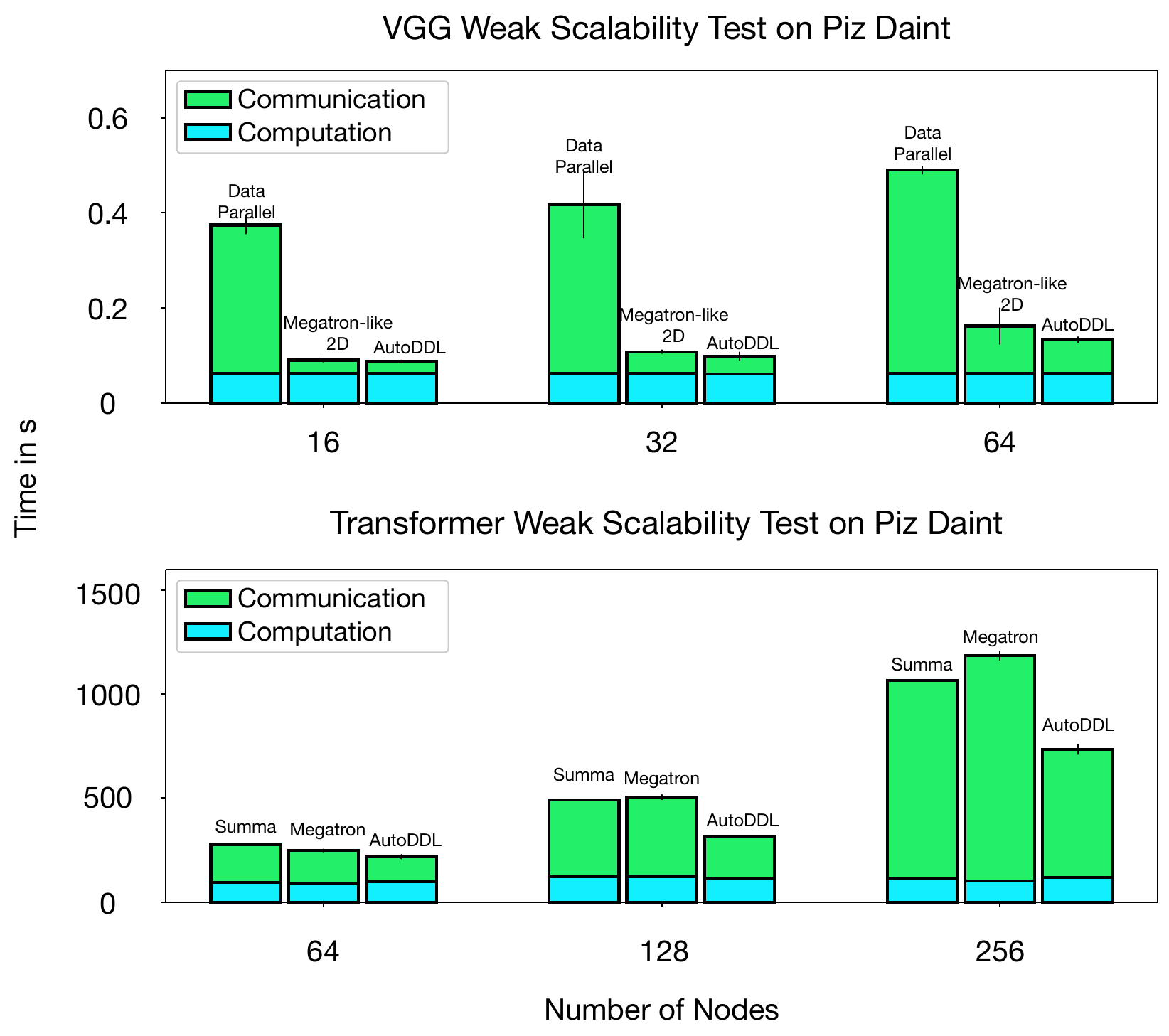}
    \vspace{-5pt}
\caption{Weak scalability test on the Piz Daint Supercomputer, which has one GPU per node. The error bars show 99\% confidence around the mean. \label{fig:daint_experiments} \vspace{-14pt}}
\end{figure}

\vspace{4pt}\noindent\textbf{VGG Results.}
The upper half of Figure~\ref{fig:daint_experiments} presents the experimental results for the VGG13 model on the Piz Daint Supercomputer using 16, 32, and 64 nodes, with batch sizes set to eight times the number of nodes. Traditional data parallelism is approximately four times slower than the strategy selected by AutoDDL. A Megatron-like 2D parallel algorithm is applied to VGG's dense layers as a stronger baseline than data parallelism. The 2D mesh topology of processes for the Megatron-like algorithm is manually tuned to achieve the best performance. For all node scales, AutoDDL consistently outperforms the Megatron-like algorithm. The performance gap increases with the number of nodes and batch sizes. Compared with the Megatron-like algorithm on 16, 32, and 64 nodes, AutoDDL reduces the communication time by 10\%, 18\%, and 29\%, respectively, and reduces the overall runtime by 3\%, 8.8\%, and 17.7\%, respectively.

\vspace{4pt}\noindent\textbf{Transformer Results.}
The lower half of Figure \ref{fig:daint_experiments} shows the performance results for Transformer model on the Piz Daint supercomputer. The batch size is fixed to be 1,024 sequences, and each has a length of 1,024 tokens. The number of attention heads is 64 for every layer, and the size of the head dimension is 128. The size of the model dimension is 8,192. We run the model on 64, 128, and 256 nodes with 8, 20, and 44 layers. The number of layers does not increase proportionally with the number of nodes because the input embedding and output logits consume a notable portion of resources and do not change with the number of layers. We set the number of layers so that the compute part is approximately the same on each GPU. The numbers of layers translate to 6.8 billion, 16.5 billion, and 35.8 billion parameters for the models. 

We manually optimize the 2D mesh topology of processes for Summa and Megatron. Data parallelism is not possible for Transformer models due to Out-of-Memory (OOM). Megatron performs better than Summa with 64 nodes, but Summa catches up Megatron when the number of nodes is larger than 128. We attribute it to the fact that Summa can alleviate the memory issues better: As the model gets larger and the number of nodes increases, the trade-off between memory and communication becomes more important. Under these circumstances, Summa saves more memory and can hence outperform Megatron. The strategy selected by AutoDDL performs the best in all cases. Compared with the second-best baseline on 64, 128, and 256 GPUs, AutoDDL reduces communication by 23.9\%, 46.4\%, and 35.3\%, respectively, and reduces the end-to-end runtime by 12\%, 36.1\%, and 31.1\%, respectively.

\subsection{Results on a Multi-Node Multi-GPU Machine}

\vspace{4pt}\noindent\textbf{System Configurations.} The experiments are conducted on two DGX A100 servers on Google Cloud, each equipped with 16 NVIDIA A100 GPUs with 40 GB of memory. The 16 GPUs within each server were connected via NvLink, providing a peak bandwidth of 600 GB/s. The servers are located in the same compute zone on Google Cloud, with a peak bandwidth of 100 Gb/s. To leverage the performance benefits of the A100's tensor cores, the neural networks are trained in mixed precision. The NCCL library is used for communication by default. Additionally, AutoDDL builds a communication model for Google Cloud and selects the optimal strategies for different models, as described in Section~\ref{sec:comm} and Section~\ref{sec:sbp_sec}. 

\vspacegraph
\begin{figure}[t]
\centering

\includegraphics[width=0.48\textwidth]{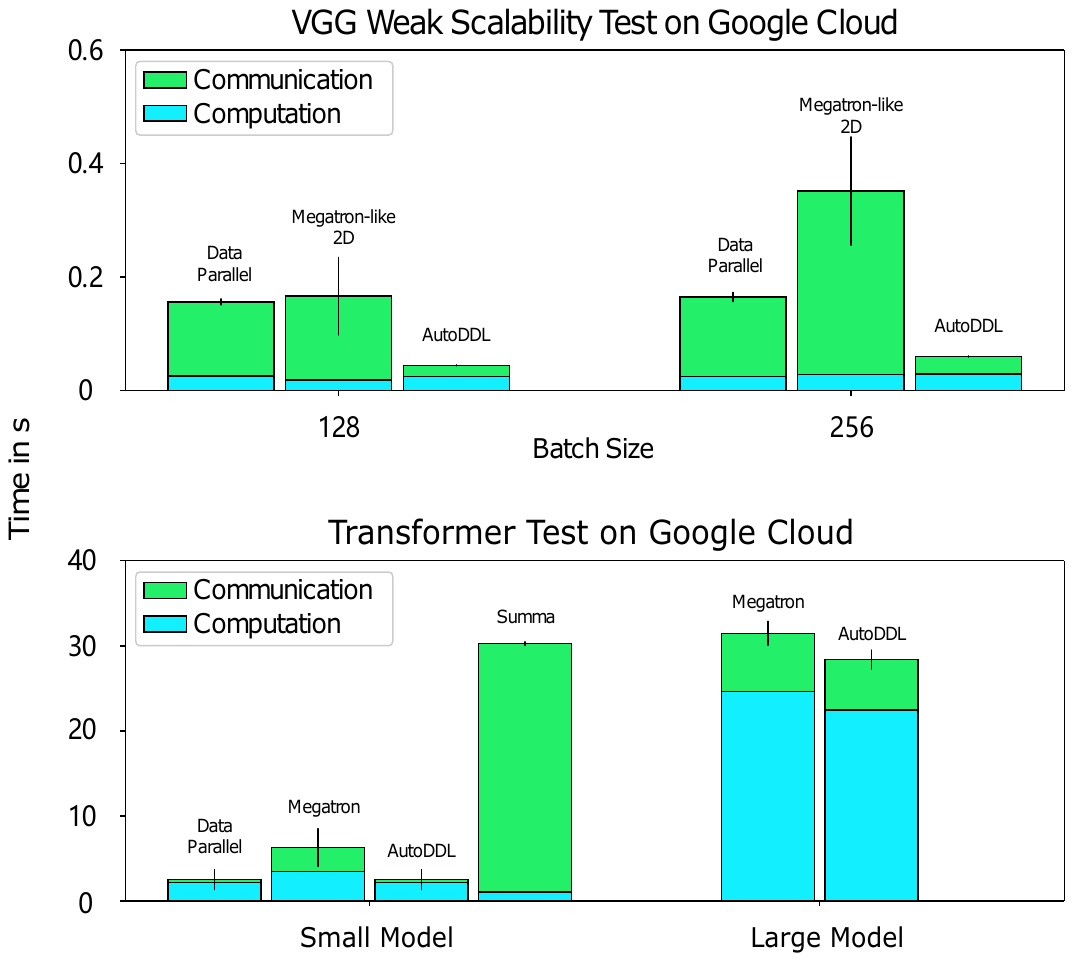}

\vspace{-5pt}
\caption{ Experiments on the Google Cloud with multiple GPUs per node. Data parallelism can't be applied to the large Transformer model due to OOM. The error bars show 99\% confidence around the mean. \label{fig:gcp_experiments} \vspace{-8pt}}
\end{figure} 

\vspace{4pt}\noindent\textbf{VGG Results.} Figure \ref{fig:gcp_experiments} shows the weak scaling results of VGG13 experiments on the Google Cloud Platform. For batch sizes of 128 and 256, 4 GPUs and 8 GPUs are used per compute node, respectively. The Megatron-like algorithm performs the worst and is unstable for both settings. AutoDDL outperforms the second-best baseline significantly: The communication is reduced by 84.7\% and 77.7\% respectively, which leads to a reduction of the overall runtime by 71.5\% and 63.5\% respectively. After analyzing the strategy selected by AutoDDL, we find that it achieves a significant improvement by 1) avoiding the expensive communication of the model weight gradients and 2) placing a large portion of communication on the fast NvLink. 

\vspace{4pt}\noindent\textbf{Transformer Results.}
Figure \ref{fig:gcp_experiments} presents the performance results for Transformer models on the Google Cloud Platform. We use all the 32 A100 GPUs in the evaluation. We test two models. The small model has 24 layers with model dimension 1,024, and the attention operation has 16 attention heads of dimension 64. The large model has 28 layers with model dimension 4,096, and the attention operation has 32 attention heads of dimension 128. The batch size is 2 million tokens for both settings: 2,048 sequences with 1,024 tokens per sequence.

The small model, which has 0.35 billion parameters, can be fitted into a single GPU memory, and data parallelism is found to be the best parallel strategy compared to other baselines. In this case, AutoDDL selects data parallelism as the optimal strategy. However, for the large model, which has 5.8 billion parameters and cannot be parallelized using data parallelism due to the OOM issue, AutoDDL selects a strategy that reduces communication by 21.6\%, resulting in 10\% runtime reduction compared with Megatron. Note that Summa parallelism for the large model is about ten times slower than AutoDDL, which is not plotted in Figure \ref{fig:gcp_experiments} for clarity.

\subsection{Pipeline Parallelism}

In this section, we demonstrate that AutoDDL offers performance advantages over hand-tuned Megatron with pipeline parallelism. Pipeline parallelism is a technique mainly used for training DNNs with deep repeated structures, and non-regular models such as CNNs (for example, VGG) are not well-suited for this approach. Therefore we only use Transformer models for the evaluation of pipeline parallelism. Megatron combines data-, operator-, and pipeline- parallelisms for Transformer models. For AutoDDL, we adopt the 1F1B pipeline schedule~\cite{pipedream,2bw}, which is widely used in deep learning frameworks such as Colossal-AI~\cite{ColossalAI} and Alpa~\cite{alpa}. Summa is not included in the baselines for two reasons. First, Summa is inherently expensive when combined with pipeline parallelism because it splits the model weights in a 2D way without any weights replication. This means that Summa has to communicate the entire model for each micro-batch, whereas AutoDDL and Megatron only communicate the entire model gradients when a mini-batch is complete. Second, unlike the other two methods, Summa requires hand-tuning of the number of micro-batches for each number of pipeline stages. Smaller micro-batch sizes imply better pipeline utilization but also more model weight communication. In addition, the 2D mesh topology of processes also needs to be tuned. Consequently, finding the best pipelined Summa implementation requires significantly more computing power, and Summa with pipeline parallelism has resulted in much worse performance compared with Megatron and AutoDDL during our experiments.

We evaluate the performance of two Transformer models (wide and deep) with different model sizes. The wide model consists of 20 layers with a model dimension of 8,192, and 64 attention heads with a head dimension of 128. On the other hand, the deep model consists of 36 layers with a model dimension of 4,096, and 32 attention heads with a head dimension of 128. 
For all baselines, we test the two models with different numbers of pipeline stages on Piz Daint with the best empirical micro-batch size for every configuration, and only report the best performance for each model. For Megatron, the configuration of operator and data parallelisms are hand-tuned to achieve the best performance. For AutoDDL, the parallelization strategy for operator and data parallelisms are automatically selected using the heuristic search algorithm.

\begin{table}[t]
\makebox[0.49 \textwidth][c]{ 
  \begin{tabular}{@{\raggedleft}|c|c|c|c|c|@{}}
  \toprule
  \hline
  Model & Scheme & Nodes & Opt. Stages & Time \\
  \hline
  \hline
  
    \multirow{2.5}{*}{Transformer} & Megatron & \multirow{2}{*}{64} & \multirow{4}{*}{1} & 533s \\
\multirow{2.5}{*}{Wide} & AutoDDL & & & 335s\\ \cline{2-3} \cline{5-5}
 & Megatron & \multirow{2}{*}{128} & & 503s \\
 & AutoDDL &  & & 315s\\ \hline
 
 \multirow{2.5}{*}{Transformer} & Megatron & \multirow{2}{*}{64} & \multirow{4}{*}{2} & 710s \\
 \multirow{2.5}{*}{Deep} & AutoDDL & & & 443s\\ \cline{2-3} \cline{5-5}
 & Megatron & \multirow{2}{*}{128} & & 807s \\
 & AutoDDL &  & & 508s\\
 \hline

 \hline
 
 \bottomrule
  \end{tabular}
  }
  \vspace{-5pt}
  \captionsetup{width=.49\textwidth}
  \caption{Performance results for Megatron and AutoDDL with pipeline parallelism enabled on Piz Daint. The numbers of optimal pipeline stages are denoted by Opt. Stage, which happen to be the same for AutoDDL and Megatron for both experiments. \vspace{-8pt}}
  \label{pipline_experiments}
\end{table}

Table~\ref{pipline_experiments} presents the results on the Piz Daint Supercomputer. On 64 nodes and 128 nodes, the batch sizes are set to 512 and 1,024 sequences, respectively, and each sequence contains 1,024 tokens. We haven't been able to scale pipeline parallelism further in our experiments. This difficulty arises potentially because more pipeline stages introduce increased overheads in terms of workload imbalance, idleness and bubbles, less compute intensity using micro-batching, point-to-point communication etc. However, when it comes to communication costs, pipeline parallelism still benefits from less costly point-to-point communications for transferring intermediate results between stages. In contrast, operator parallelism relies on more expensive collective operations to collect intermediate results for each layer. And AutoDDL is able to discover a near-optimal 3D parallelization strategy for operator and data parallelism, which significantly alleviates the communication bottleneck, and therefore outperforms the hand-tuned Megatron with pipeline parallelism in all of the experiments as listed in Table~\ref{pipline_experiments}.


\section{Conclusions}
AutoDDL automatically discovers and implements parallelization strategies with near-optimal bandwidth cost for distributed deep learning. Based on the abstraction of SBP, we created a broader range of parallelization strategies, making it practical to find communication-optimal approaches that genuinely improve performance. Furthermore, we equip AutoDDL with a heuristic search algorithm (coordinate descent) based on performance modelling to quickly find near-optimal strategies in the extended search space. We integrate OneFlow into the workflow of AutoDDL, which enables automatic implementation of various parallelization strategies and thus achieves high productivity. We conduct the evaluation on the Piz Daint Supercomputer and a Google Cloud A100 GPU platform using different neural networks (VGG and GPT). Experimental results show that the parallelization strategy selected and realized by AutoDDL outperforms all the baselines (including the highly optimized schemes by hand-tuning) on different distributed machines. 

This work brings new insights for the optimization of distributed deep learning. Firstly, exploring near-optimal strategies at a per-layer/operator granularity is essential to improve the overall performance of the neural network. Because the training processes may have different batch sizes and layer structures, each layer may require a distinct parallelization strategy to achieve high performance. For example, Figure \ref{fig:strategy_example} shows how different parallelization strategies are selected for the last three dense layers of VGG13. Secondly, AutoDDL's expanded search space enables practical use of asymptotically optimal parallelization strategies, leading to the discovery of more effective execution plans for actual performance improvements. The introduction of the Partial-Sum tensor state in AutoDDL broadens the search space, allowing for its transformation into Split states through an efficient ReduceScatter. Previous deep learning frameworks, such as Alpa and Flexflow, are built upon the GShard abstraction with limited search spaces. Benefiting from the expanded space, AutoDDL significantly outperforms expert-optimized Megatron in real experiments, while previous frameworks cannot. Lastly, the coordinate descent algorithm based on performance modelling used in AutoDDL significantly accelerates the search process. Experimental results show that AutoDDL can find near-optimal strategies in minutes using a personal laptop. A fast search algorithm with low resource requirement greatly improves the practicability of the automatic deep learning framework, which is more desirable for training more complex models on future parallel systems.

\section{Acknowledgement}
This project was supported by the National Natural Science Foundation of China under Grant No. 62372055, the Sci-Tech Innovation 2030 Agenda of China (2023ZD0120502), the Fundamental Research Funds for the Central Universities, the PASC DaCeMI project, also received funding from the European
High-Performance Computing Joint Undertaking (JU) under grant agreement No. 101034126 (EU-Pilot). We thank CSCS for providing us access to compute resources.

\bibliographystyle{IEEEtran}
\bibliography{mybib}




\end{document}